\documentclass[%
 preprint,
 amsmath,amssymb,
 aps,prmaterials,
]{revtex4-2}
\usepackage{graphicx}
\usepackage{dcolumn}
\usepackage{bm}
\begin{document}
\preprint{APS/123-QED}
\title{DSpinGNN: A Physics-Informed Equivariant Graph Neural Network for Dynamic
Magnetic Exchange Prediction in Strain-Deformed Monolayer CrI$_3$}
\author{Isam Abdullah Balghari}%
 \email{isamabdullah88@gmail.com}
 \affiliation{%
 Department of Physics, Syed Babar Ali School of Science and Engineering,
 Lahore University of Management Sciences (LUMS), Opposite Sector U,
 DHA, Lahore 54972, Pakistan.
 }
\author{Muhammad Faryad}%
 \email{muhammad.faryad@lums.edu.pk}
\affiliation{%
 Department of Physics, Syed Babar Ali School of Science and Engineering,
 Lahore University of Management Sciences (LUMS), Opposite Sector U,
 DHA, Lahore 54972, Pakistan.
}
\author{Muhammad Sabieh Anwar}%
 \email{sabieh@lums.edu.pk}
\affiliation{%
 Department of Physics, Syed Babar Ali School of Science and Engineering,
 Lahore University of Management Sciences (LUMS), Opposite Sector U,
 DHA, Lahore 54972, Pakistan.
}%
\date{\today}
\begin{abstract}
Resolving the instantaneous, position-dependent isotropic magnetic exchange coupling
$J_{ij}$ across a dynamically deforming crystal lattice requires a computational
approach that simultaneously handles structural forces and magnetic interactions at
length scales inaccessible to first-principles methods. Here we introduce DSpinGNN,
a bifurcated machine-learning architecture comprising an $E(3)$-equivariant graph neural
network (E-GNN) for classical Langevin structural dynamics and a physics-informed
$\Delta$-MLP that maps instantaneous local Cr-I-Cr bond geometry to isotropic exchange
couplings, with the Goodenough-Kanamori superexchange relationship embedded as an
analytical inductive bias. Trained on 345 DFT+U configurations of monolayer CrI$_3$
and evaluated on a strictly withheld 61-configuration test set, DSpinGNN simultaneously
achieves an energy MAE of 1.1~meV/atom, a force MAE of 6.5~meV/\AA, and an exchange
coupling MAE of 0.18~meV ($R^2 = 0.91$). Deployed at 400$\times$ scale in a 3,200-atom
supercell under a collinear Ising-constrained adiabatic approximation at 5~K, the model
maps the local exchange response to a propagating biaxial strain wave. Wave reflection at
periodic boundaries generates transient constructive interference regions where local
compressive strain exceeds the DFT-established FM-to-AFM threshold, producing spatially
heterogeneous exchange coupling textures that damp as the wave dissipates. Quantitative
analysis yields a domain wall width of $\xi = 1.7 \pm 0.3$~nm and a
constructive-interference oscillation period of $\tau = 0.27$~ps---mesoscopic observables
inaccessible to direct DFT and constituting testable predictions for cryogenic magnetic
force microscopy. DSpinGNN provides a reproducible, transferable framework for mesoscale
exchange mapping in strain-driven 2D magnetic materials.
\end{abstract}
\maketitle
\section{Introduction}
The discovery of intrinsic long-range ferromagnetic order in monolayer CrI$_3$ at
cryogenic temperatures~\cite{Huang2017_Nature} established that magnetic order can
survive at the strict two-dimensional limit of van der Waals materials, opening new
directions for investigating fundamental magnetic physics in reduced dimensions. As the
prototypical 2D van der Waals magnetic insulator, monolayer CrI$_3$ exhibits a
ferromagnetic ground state whose stability is intimately coupled to its structural
geometry: the competition between ferromagnetic (FM) superexchange and antiferromagnetic
(AFM) direct exchange is governed by the Cr-I-Cr bridging angles and Cr-Cr interatomic
distances, as dictated by the Goodenough-Kanamori (GK)
rules~\cite{Goodenough1955_PhysRev, Kanamori1959_JPCS, Webster2018_PRB, Jiang2018_PRB,
McGuire2017_Crystals}. Consequently, applied lattice strain---which modifies both bond
angles and bond lengths---offers a non-volatile mechanism for tuning the magnetic ground
state without external magnetic fields~\cite{Wang2020_AdvMat}.
A key physical constraint governs all computational treatments of 2D magnetism and must
be stated at the outset. In a strictly two-dimensional system with isotropic (Heisenberg)
spin interactions, the Mermin-Wagner theorem prohibits spontaneous long-range magnetic
order at any finite temperature~\cite{Mermin1966_PRL}. The experimentally observed
ferromagnetic order in monolayer CrI$_3$ survives not because of isotropic exchange
alone, but because spin-orbit coupling (SOC) generates a finite single-ion magnetic
anisotropy (SIA) and anisotropic exchange terms---including Kitaev-type contributions
mediated by the heavy iodine ligands---that break the continuous spin-rotation symmetry
and lift the Mermin-Wagner prohibition~\cite{Lee2020_PRL, Lado2017_2DMat}. This has a
direct implication for computational modeling: any framework that omits SOC from its
training data, as the present work does, must impose an explicit symmetry-breaking
constraint in its place to maintain a well-defined collinear ground state. We address this
directly in the Methods section.
The sensitivity of $J_{ij}$ to structural geometry motivates strain engineering as a
pathway for continuous magnetic tuning. Biaxial compressive strain uniformly reduces
Cr-Cr distances, exponentially enhancing direct $d$-$d$ orbital overlap and driving a
FM-to-AFM transition near $-5\%$ to $-6\%$ compression~\cite{Zheng2018_Nanoscale,
Webster2018_PRB, Sivadas2018_NanoLett}. The physical relevance of mesoscale structural
heterogeneity is underscored by experiments: cryogenic magnetic force microscopy has
revealed coexisting FM and AFM domains in CrI$_3$ flakes attributed to local stacking
faults~\cite{Thiel2019_Science}, spatially resolved circular dichroism measurements have
correlated structural domain boundaries with adjacent magnetic phase
regions~\cite{Chen2019_Science}, and hydrostatic pressure experiments have demonstrated
continuous mechanical tuning between mixed magnetic states~\cite{Song2019_NatMat}.
Together, these observations motivate a computational framework capable of resolving the
local magnetic exchange response to mesoscale structural deformations.
Simulating this dynamic magneto-structural response presents a fundamental computational
bottleneck. First-principles molecular dynamics evolves lattice and spin degrees of
freedom with quantum mechanical accuracy, but its $\mathcal{O}(N^3)$ scaling restricts
practical calculations to supercells of fewer than $\sim$100 atoms over
picoseconds~\cite{Evans2014_JPCM, Hellman2020_PRM}---prohibiting direct access to strain
wave propagation~\cite{Kuss2022_APL}, substrate corrugation~\cite{Woods2014_NatPhys}, or
meso-scale domain formation. Classical molecular dynamics can reach the required scales
but contains no representation of the quantum magnetic exchange~\cite{Plimpton1995_JCP}.
Machine learning interatomic potentials (MLIPs) have resolved the structural side of
this gap, with equivariant architectures achieving near-DFT accuracy on complex materials
at classical-MD cost~\cite{Batzner2022_NatCommun, Musaelian2023_NatCommun}. Extending
MLIPs to magnetic systems, where atomic forces and exchange interactions must be
simultaneously captured, is an active area of development~\cite{Eckhoff2021_npjCM,
Drautz2019_PRB}. Eckhoff and Behler coupled spin-polarized neural network potentials to
a Heisenberg Hamiltonian via an atom-centered descriptor framework, demonstrating dynamic
exchange prediction in bulk magnetic materials~\cite{Eckhoff2021_npjCM}. More recently,
the SpinGNN framework introduced time-reversal equivariant tensor representations to
predict full $3\times3$ exchange tensors including anisotropic and DMI
components~\cite{Yu2023_SpinGNN}. These advances notwithstanding, their application to
2D magnetic materials under continuous strain deformation remains unexplored.
Additionally, existing ML approaches to 2D magnetism have predominantly targeted static
binary classification (FM vs.\ AFM)~\cite{Rhone2020_AdvMat}, leaving a gap for models
that predict continuous $J_{ij}$ values dynamically across thousands of atoms during
time-resolved structural evolution.
DSpinGNN addresses this gap with a targeted design: an $E(3)$-equivariant GNN for
force-driven structural dynamics, coupled adiabatically to a physics-informed
$\Delta$-MLP exchange predictor that embeds the Goodenough-Kanamori relationship as an
inductive bias~\cite{Pizzochero2020_JPCC, Kashin2020_2DMat}. The framework is explicitly
designed for the collinear isotropic limit, which captures the dominant FM-to-AFM
competition under strain while remaining tractable for mesoscale simulation. By training
exclusively on 8-atom primitive cells and deploying on a 3,200-atom supercell, we
demonstrate the length-scale transferability of $E(3)$-equivariant architectures in a
magnetic materials context and extract quantitative mesoscopic observables---domain wall
widths and oscillation timescales---that constitute testable predictions inaccessible to
direct DFT.
\begin{figure*}[t]
    \centering
    \includegraphics[width=\textwidth]{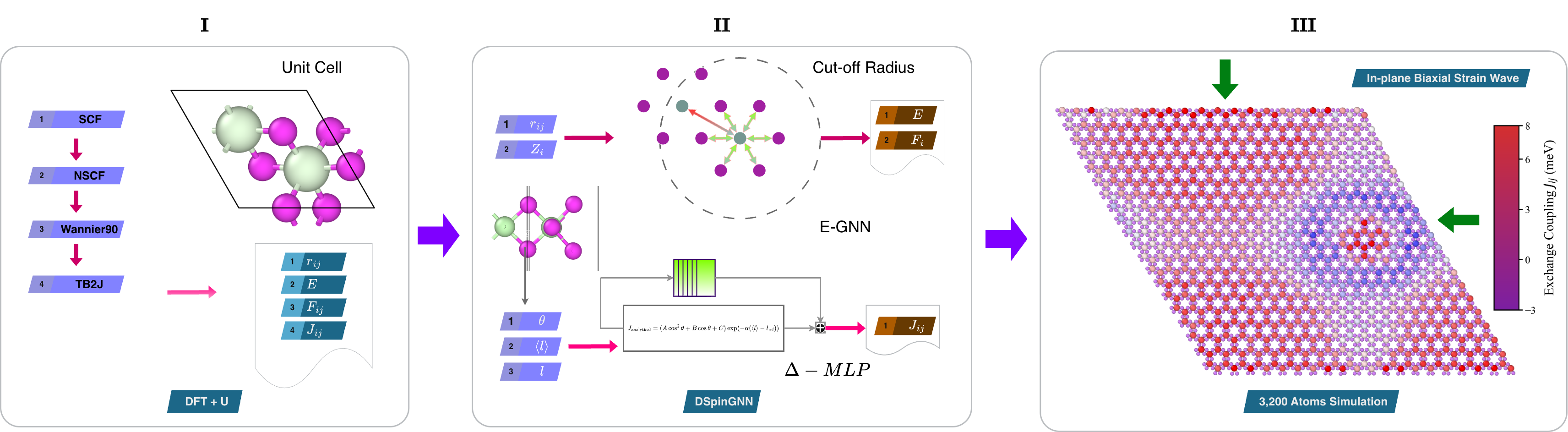}
    \caption{\label{fig:pipeline}
    Complete workflow of the DSpinGNN framework for mesoscale dynamic magnetic mapping.
    \textbf{(I) Data Generation:} A DFT+U pipeline (Quantum ESPRESSO $\to$ pw2wannier90
    $\to$ Wannier90 $\to$ TB2J) maps structural configurations to total energies ($E$),
    atomic forces ($F_i$), and isotropic first-nearest-neighbor exchange couplings
    ($J_{ij}$).
    \textbf{(II) Model Architecture:} The bifurcated architecture separates predictions
    by physical content. The E-GNN branch predicts scalar energy and equivariant atomic
    forces from species and edge vectors via $E(3)$-equivariant message passing. The
    independent $\Delta$-MLP processes Cr-I-Cr local geometry (bridging angles $\theta$,
    bond lengths $l$) with an analytical Goodenough-Kanamori block providing a
    physics-informed baseline for the isotropic exchange prediction. Under the adiabatic
    approximation, the E-GNN drives the structural trajectory while the $\Delta$-MLP
    maps instantaneous bond geometry to exchange couplings as a post-hoc operation.
    \textbf{(III) Mesoscale Simulation:} DSpinGNN drives 3,200-atom Langevin dynamics.
    Reflection of the biaxial strain wave at periodic boundaries and subsequent
    constructive superposition generate transient high-strain regions, enabling mapping
    of strain-induced exchange heterogeneity at mesoscopic length scales.
    }
\end{figure*}
\section{Computational Methods}
\subsection{First-Principles Data Generation}
A comprehensive illustration of the full pipeline is shown in Fig.~\ref{fig:pipeline}.
First-principles DFT calculations were performed using the Quantum ESPRESSO
package~\cite{Giannozzi2009_JPCM, Giannozzi2017_JPCM}. The exchange-correlation
functional was treated within GGA-PBE~\cite{Perdew1996_PRL}. To capture the localized
nature of the Cr $3d$-orbitals, a Hubbard correction of $U = 3.0$~eV was
applied~\cite{Dudarev1998_PRB}. This value is consistent with the range commonly employed
for CrI$_3$ in the literature (2.0--4.0~eV)~\cite{Webster2018_PRB, Zheng2018_Nanoscale}.
Grimme DFT-D3 van der Waals corrections~\cite{Grimme2010_JCP} were applied to accurately
resolve in-plane lattice constants and the sensitive Cr-I-Cr superexchange angles. Core
electrons were described using the SSSP Efficiency (v1.3.0) pseudopotential
library~\cite{Prandini2018_npjCM} with an energy convergence threshold of $10^{-9}$~Ry.
The structural dataset was built using the 8-atom primitive unit cell of monolayer
CrI$_3$. The pristine cell was fully relaxed (variable-cell) on an $8\times8\times1$
Monkhorst-Pack $k$-point mesh~\cite{Monkhorst1976_PRB}. Biaxial, uniaxial ($x$-axis),
and shear ($xy$-plane) strains were applied in the range $-5\%$ to $+5\%$ in 11 discrete
steps. Each strained configuration was structurally relaxed before atomic rattling with
Gaussian displacements ($\sigma = 0.02$ and $0.04$~\AA), yielding 3--5 configurations
per strain state. For each configuration, self-consistent field (SCF) and
non-self-consistent field (NSCF) calculations (55 bands) were performed on an
$8\times8\times1$ $k$-mesh. The \texttt{pw2wannier90} interface was used to project the
Bloch states onto maximally localized Wannier functions generated by
\texttt{Wannier90}~\cite{Pizzi2020_JPCM}, yielding a real-space tight-binding
Hamiltonian. Isotropic magnetic exchange couplings were subsequently extracted using the
\texttt{TB2J} code~\cite{He2021_PhysicaB} via the Liechtenstein magnetic force
theorem~\cite{Liechtenstein1987_JMMM} on a dense $36\times36\times1$ $k$-mesh for
numerical convergence of the exchange integrals.
Extended DFT energy curves (up to $\pm15\%$ strain) were computed under constrained
structural relaxation for phase boundary context. Full TB2J downfolding across this
extended range was computationally prohibitive and was not performed; these extended
curves serve only as qualitative context for the energy landscape and are presented in
the Supplemental Material~\cite{SupplementalMaterial} (Figs.~S1--S3).
\subsection{Magnetic Hamiltonian and Explicit Physical Approximations}
The magnetic interactions are mapped onto the classical Heisenberg Hamiltonian:
\begin{equation}
    \mathcal{H}_{\text{spin}} = -\sum_{\langle i,j \rangle} J_{ij}(\mathbf{r})\,
    \mathbf{S}_i \cdot \mathbf{S}_j
    \label{eq:heisenberg}
\end{equation}
where $J_{ij}(\mathbf{r})$ is the position-dependent isotropic exchange scalar
restricted to first-nearest-neighbor Cr-Cr pairs. This framework rests on three explicit
approximations that define the physical scope of all results:
\textit{Approximation 1: Collinear Ising constraint.} Spin vectors are confined to an
Ising-like easy axis (parallel or antiparallel along $\hat{z}$). This constraint serves
as a minimal proxy for the spin-orbit-coupling-driven single-ion anisotropy that
physically stabilizes the collinear magnetic state in CrI$_3$ against thermal
fluctuations and Mermin-Wagner divergences at finite temperature. Since SOC is omitted
from the DFT dataset (see Approximation 2), this imposed constraint is necessary to
maintain a well-defined collinear ground state during the simulation. Quantities
requiring continuous spin-rotation freedom---magnon spectra, domain wall spin textures,
non-collinear order---are outside the scope of this model.
\textit{Approximation 2: Collinear DFT; omission of SOC.} This study targets the binary
FM/AFM competition in monolayer CrI$_3$ under strain, governed by the isotropic
Heisenberg exchange $J_1$. The sign change of $J_1$ from positive (FM) to negative (AFM)
is driven by the Goodenough-Kanamori competition between orthogonal superexchange and
direct $d$-$d$ overlap---a purely isotropic exchange mechanism that is independent of
SOC. SOC-driven terms in CrI$_3$---single-ion anisotropy, Kitaev-type exchange, and the
Dzyaloshinskii-Moriya interaction~\cite{Lee2020_PRL, Lado2017_2DMat}---are corrections
to the anisotropic part of the Hamiltonian and do not govern the sign or magnitude of
the isotropic $J_1$. The collinear DFT treatment with the Liechtenstein force theorem as
implemented in TB2J~\cite{He2021_PhysicaB} is therefore sufficient for the restricted
objective of this work, and has been validated against non-collinear treatments in the
collinear-dominant strain regime~\cite{Drchal2013_EPJWC}. Second- and third-neighbor
exchange couplings ($J_2$, $J_3$), which influence finer features of the magnetic ground
state~\cite{Lado2017_2DMat}, are not included; their omission is justified by the goal
of capturing the primary FM-to-AFM transition driven by $J_1$.
\textit{Approximation 3: Adiabatic decoupled approximation.} The structural E-GNN and
magnetic $\Delta$-MLP branches operate in a decoupled adiabatic regime. Langevin
dynamics are driven exclusively by the E-GNN force predictions; no magnetic energy
contribution enters the force calculation. The $\Delta$-MLP maps instantaneous bond
geometry---output at each time step by the structural trajectory---to exchange coupling
values as a post-hoc operation. This decoupling rests on the adiabatic approximation
of spin dynamics~\cite{Antropov1995_PRL, Streib2020_PRB}: the isotropic exchange
coupling $J_{ij}(\mathbf{r})$ is an electronic-structure quantity that adjusts to the
instantaneous atomic geometry on femtosecond timescales, far faster than the phonon
dynamics that drive structural evolution. GHz-frequency spin resonances observed in
related Cr trihalide materials~\cite{MacNeill2019_PRB} confirm that spin dynamics in
this class of magnets operate on picosecond-to-sub-nanosecond timescales, consistent
with treating $J_{ij}$ as an adiabatic function of the nuclear coordinates. An
important consequence is that the framework does not capture the back-action of the
magnetic state on phonon dynamics---a higher-order magneto-elastic effect. The exchange
coupling maps in the Results section therefore represent the predicted \textit{adiabatic
magnetic exchange landscape} of the deforming lattice.
The competition between FM and AFM exchange is governed by two competing pathways. The
primary FM contribution arises from superexchange: virtual hopping of electrons between
the half-filled Cr $t_{2g}$ and empty Cr $e_g$ orbitals via the bridging I $5p$ ligands.
Because the pristine Cr-I-Cr bond angle is near $90^\circ$, the GK rules dictate that
this orthogonal pathway yields ferromagnetic alignment~\cite{Goodenough1955_PhysRev,
Kanamori1959_JPCS}. Under compressive strain, reduced Cr-Cr distances force direct
$d$-$d$ orbital overlap; the Pauli exclusion principle then drives antiferromagnetic
alignment~\cite{Kashin2020_2DMat, Khomskii2014_Book}. Because this balance is sensitive
to picometer-scale lattice distortions, a static $J$ in molecular dynamics is
insufficient, and dynamic prediction of $J_{ij}(\mathbf{r})$ forms the primary objective
of the $\Delta$-MLP.
\subsection{DSpinGNN Architecture and Training}
The structural branch employs the NequIP E-GNN framework~\cite{Batzner2022_NatCommun}
with 3 interaction layers and a $7.0$~\AA\ cutoff. $E(3)$ equivariance enforced
throughout message passing~\cite{Geiger2022_e3nn, Thomas2018_TFN} ensures scalar outputs
are invariant and forces transform correctly under all rigid-body transformations,
eliminating rotational data augmentation and providing the length-scale transferability
demonstrated at deployment.
The $\Delta$-MLP exchange predictor processes Cr-I-Cr local subgraphs using the bridging
angle $\theta$, the four Cr-I leg lengths $l_1, l_2, l_3, l_4$ (mean: $\langle l
\rangle$), and the Cr-Cr interatomic distance. The physics-informed analytical block
implements the Goodenough-Kanamori-inspired ansatz using $\theta$ and $\langle l \rangle$:
\begin{equation}
  J_{\text{analytical}} = \left(A\cos^2\theta + B\cos\theta + C\right)
  \exp\!\left(-\alpha\!\left(\langle l\rangle - l_{\mathrm{ref}}\right)\right)
  \label{eq:gk_ansatz}
\end{equation}
where $l_{\mathrm{ref}} = 2.70$~\AA\ is the equilibrium Cr-I bond length
in the fully relaxed pristine cell. The parameters $A$, $B$, $C$, and $\alpha$ are
treated as learnable parameters initialized with physically motivated estimates and
optimized jointly with the residual MLP during end-to-end training. Their converged
values, which encode the GK-inspired angular and bond-length dependence of the exchange
coupling, are reported in Table~\ref{tab:gkparams}. The full output is:
\begin{equation}
J_{ij} = J_{\text{analytical}}(\theta, \langle l\rangle) + J_{\text{residual}}(\mathbf{h})
\end{equation}
where $J_{\text{residual}}$ is a shallow MLP operating on the local geometric feature
vector $\mathbf{h}$. In regions of large deformation, the residual contribution
diminishes relative to the analytical baseline, providing asymptotic stability against
unphysical divergence in extrapolation.
Full training details and hyperparameters are provided in the Appendix.
\subsection{Large-Scale Structural Dynamics and Local Strain Analysis}
Mesoscale simulations were conducted using ASE~\cite{Larsen2017_JPCM}. A $20\times20$
CrI$_3$ supercell (3,200 atoms, dimensions $140.1$~\AA~$\times$~$121.3$~\AA) was relaxed using
FIRE~\cite{Bitzek2006_PRL} to a force tolerance of $10^{-4}$~eV/\AA\ prior to
dynamical integration. Langevin dynamics were integrated with a time step of 5~fs and a
damping constant $\gamma = 0.982$~ps$^{-1}$~\cite{Allen2017_Book} at
$T = 5$~K. The low temperature suppresses thermal phonon noise while retaining
sufficient configurational flexibility for strain-driven structural dynamics.
A biaxial sinusoidal displacement field:
\begin{equation}
u_\alpha(\mathbf{r}_0) = A_0 \sin\!\left(\frac{2\pi r_{0,\alpha}}{\lambda_\alpha}\right),
\quad \alpha \in \{x, y\}
\end{equation}
with amplitude $A_0 = 1.5$~\AA\ and wavelength $\lambda_\alpha = 4L_\alpha$ for each
in-plane direction $\alpha \in \{x,y\}$, was applied to the initial atomic coordinates
and the system was subsequently allowed to relax under Langevin dynamics. The
corresponding maximum nominal sinusoidal strains are
$\varepsilon_{\rm nom}^x = 2\pi A_0/\lambda_x \approx 1.7\%$ and
$\varepsilon_{\rm nom}^y = 2\pi A_0/\lambda_y \approx 1.9\%$, both well below the
DFT-established FM-to-AFM threshold of $\approx -6\%$.
To quantify the local deformation, the local strain tensor was computed at each time step
from the deformation gradient tensor $\mathbf{F}$. For each atom $i$, $\mathbf{F}$ is
evaluated by comparing the current positions of its neighbors to their reference
positions in the relaxed lattice, using the Voronoi tessellation to define each atom's
local coordination volume (i.e., the set of all points in space closer to atom $i$ than
to any other atom)~\cite{Falk1998_PRE, Stukowski2010_MSMSE}:
\begin{equation}
\varepsilon_{\rm local} = \frac{1}{2}\!\left(\mathbf{F} + \mathbf{F}^T\right) - \mathbf{I}
\end{equation}
The scalar local compressive strain $|\varepsilon_{\rm local}|$ is the magnitude of the
most compressive principal strain eigenvalue of $\varepsilon$, providing a per-atom
measure for direct comparison with the DFT phase boundary.
\section{Results and Discussion}
\subsection{Strain-Induced Magnetic Phase Boundaries: DFT Reference}
DFT total energy calculations for monolayer CrI$_3$ under biaxial, uniaxial, and shear
deformations confirm the strain-dependent magnetic phase landscape (Supplemental
Figs.~S1--S3~\cite{SupplementalMaterial}). Biaxial compression drives a FM-to-AFM transition at approximately
$-6\%$ strain, consistent with prior calculations~\cite{Zheng2018_Nanoscale,
Webster2018_PRB}, as the uniform reduction in Cr-Cr distances enhances direct $d$-$d$
orbital overlap. Uniaxial compression delays the transition to approximately $-12\%$ via
the Poisson effect. Pure in-plane shear does not induce a global FM-to-AFM transition up
to $\pm15\%$ because the area-preserving deformation preserves the dominant orthogonal
superexchange pathways~\cite{Pizzochero2020_JPCC}. These results establish the biaxial
FM-to-AFM threshold ($\approx -6\%$ compressive strain) as the reference boundary for
interpreting the exchange coupling maps in the mesoscale simulation.
\subsection{Model Validation: Independent Test Set Performance}
\begin{figure}[htbp]
\includegraphics[scale=0.95]{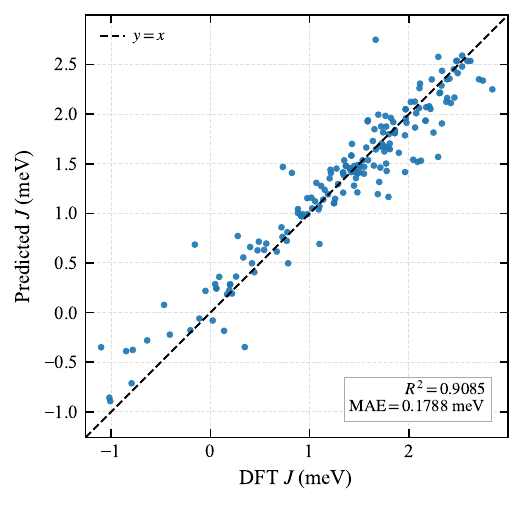}
\caption{\label{fig:jvalidation}
Parity plot of DSpinGNN predicted exchange coupling $J_{ij}$ vs.\ DFT+U ground-truth
values for all first-nearest-neighbor Cr-Cr bonds sampled from the 61 held-out test
configurations. Ground-truth exchange values are from the same Quantum ESPRESSO $\to$
Wannier90 $\to$ TB2J pipeline used to build the training data. The dashed line denotes
perfect agreement. The test set MAE of 0.18~meV and $R^2 = 0.91$
confirm that the physics-informed $\Delta$-MLP generalizes accurately to unseen
structural configurations. Only the exchange coupling is shown; energy and force
predictions on the same test set are reported in Table~\ref{tab:performance}.
}
\end{figure}

The E-GNN structural branch and $\Delta$-MLP exchange predictor were jointly evaluated
on a strictly withheld test set of 61 configurations. These configurations were
generated using the same DFT+U pipeline after all model development---including
architecture decisions, hyperparameter tuning, and early stopping---was finalized, and
were never presented to the model at any stage. Dataset splits were stratified across
deformation mode (biaxial, uniaxial, shear) and strain magnitude bin, ensuring each
split retains a representative distribution of the full deformation space
(see Appendix~\ref{sec:dataset}). The test set was evaluated exactly once.
All three predicted quantities---energy, forces, and exchange couplings---were assessed
simultaneously against their DFT+U ground-truth values, providing a unified and
unbiased measure of generalization. Performance is summarized in
Table~\ref{tab:performance}. The close agreement between validation and test MAEs
confirms the absence of overfitting to the validation split.
\begin{table}[htbp]
\caption{\label{tab:performance}
DSpinGNN performance metrics across dataset splits
(345 train / 61 validation / 61 test). The 61-configuration test set was generated
after all model development was complete, was never seen by the model during training
or selection, and was evaluated exactly once with all three quantities assessed
simultaneously against DFT+U ground truth.}
\begin{ruledtabular}
\begin{tabular}{lccc}
Quantity & Training MAE & Validation MAE & Test MAE \\
\hline
Energy (meV/atom) & 0.9  & 1.0  & 1.1  \\
Force (meV/\AA)   & 4.0  & 6.0  & 6.5  \\
$J_{ij}$ (meV)    & 0.20 & 0.27 & 0.18 \\
\end{tabular}
\end{ruledtabular}
\end{table}
To visualize the exchange coupling accuracy specifically,
Fig.~\ref{fig:jvalidation} shows a parity plot of DSpinGNN predicted $J_{ij}$ values
against their DFT+U ground-truth counterparts across all bonds sampled from the 61 test
configurations. Note that energy and force predictions were evaluated simultaneously on
the same test set but are not shown here; their MAEs are reported in
Table~\ref{tab:performance}.
\subsection{Mesoscale Exchange Coupling Heterogeneity Under Strain Wave Propagation}
\begin{figure*}[t]
    \centering
    \includegraphics[scale=0.95]{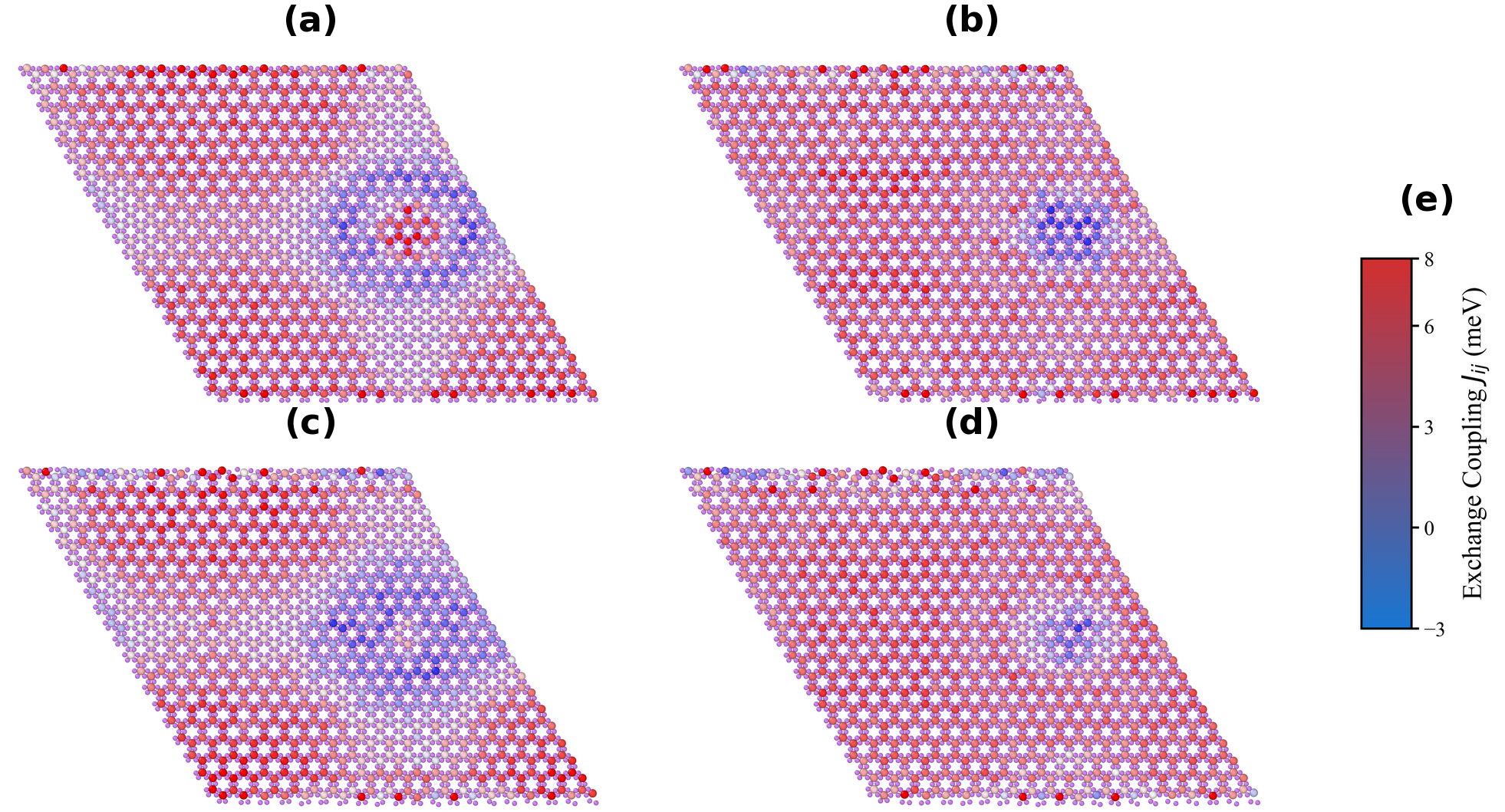}
    \caption{\label{fig:simulation}
    Predicted exchange coupling landscape during strain wave propagation and reflection
    in a 3,200-atom CrI$_3$ monolayer. Panels \textbf{(a)}-\textbf{(d)} show
    consecutive time steps following the onset of wave reflection. Red: FM-sign ($J>0$);
    blue: AFM-sign ($J<0$). Constructive superposition of incident and reflected wave
    components drives localized compressive regions beyond the $\approx-6\%$ DFT
    threshold, generating spatially heterogeneous exchange coupling zones. Panel
    \textbf{(a)}: peak AFM-sign extent; \textbf{(b)}: contraction of the AFM zone;
    \textbf{(c)}: further recovery; \textbf{(d)}: near-full FM-sign restoration as the
    wave dissipates. Colorbar for $J$ is provided in \textbf{(e)}.
    }
\end{figure*}

\begin{figure*}[t]
    \centering
    \includegraphics[scale=0.95]{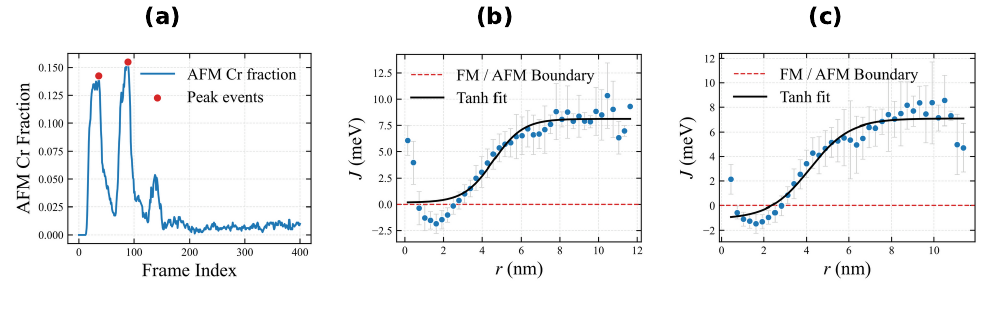}
    \caption{\label{fig:metrics}
    Quantitative mesoscale observables from the 3,200-atom DSpinGNN simulation.
    \textbf{(a)} AFM Cr fraction $\mathcal{F}(t)$ as a function of trajectory frame
    (5~fs per frame), showing two constructive interference events (red circles) at
    frames 36 and 89 arising from superposition of the incident and reflected biaxial
    strain wave. The inter-event spacing of 53 frames yields $\tau = 0.27$~ps. The
    fraction decays to a low thermal baseline after frame $\sim$150 as the wave
    dissipates through the Langevin thermostat.
    \textbf{(b)} Radial exchange coupling profile $J(r)$ at the frame~36 constructive
    interference peak. Hyperbolic tangent fit (Eq.~\ref{eq:tanh_fit}, black curve) yields
    $\xi = 1.52 \pm 0.52$~nm ($R^2 = 0.84$).
    \textbf{(c)} Radial profile at the frame~89 peak, yielding
    $\xi = 1.90 \pm 0.48$~nm ($R^2 = 0.91$). The mean over both events is
    $\xi = 1.7 \pm 0.3$~nm. In both panels, $J$ values at large $r$ reflect tensile
    strain at the interference periphery strengthening the FM superexchange pathway
    relative to equilibrium; the dotted horizontal line marks $J = 0$.
    }
\end{figure*}

We now report results from the 3,200-atom Langevin simulation. All exchange coupling
maps represent the adiabatic $J_{ij}$ landscape predicted from instantaneous bond
geometry (Sec.~II.B, Approximation~3).
\textbf{Wave reflection and local strain focusing.} The applied sinusoidal displacement
field corresponds to maximum nominal strains of $\approx 1.7\%$ ($x$-direction) and
$\approx 1.9\%$ ($y$-direction), both below the FM-to-AFM threshold.
During the initial propagation phase, no AFM-sign exchange regions are observed. As the
wave reaches the periodic boundaries of the supercell and reflects back, the superposition
of incident and returning wave components creates transient regions of constructive
interference. Deformation gradient analysis of the Cr sublattice confirms that within
these interference regions, the local compressive strain transiently exceeds $-6\%$ for
approximately 120~fs (first event, frames 17--40) and 380~fs (second event, frames
64--139), with Cr-sublattice compressive strains of $\varepsilon_{\rm Cr} = -7.1\%$
and $-9.1\%$ at the respective domain wall extraction frames (36 and 89). The time evolution of the peak Cr-sublattice strain through
both interference events, confirming the above-threshold episodes, is shown in
Supplemental Figs.~S4a--S4b~\cite{SupplementalMaterial}. The above-threshold
Cr-sublattice strain zone spans an effective radius of approximately 6~nm at peak
compression, encompassing a coherent AFM exchange domain whose spatial sharpness is
characterized by the domain wall width $\xi = 1.7 \pm 0.3$~nm.
\textbf{Spatially heterogeneous exchange coupling maps.} When localized strain exceeds
the FM-to-AFM threshold, the per-bond $J_{ij}$ map develops central zones of AFM-sign
exchange ($J_{ij} < 0$) surrounded by a FM-sign ($J_{ij} > 0$) background, with
approximately circular geometry reflecting the isotropic biaxial strain field
(Fig.~\ref{fig:simulation}). As the wave propagates and local strain relaxes, the
AFM-sign zone contracts and the FM-sign background recovers
(Fig.~\ref{fig:simulation}(a)--(d)); the cycle dampens progressively as the wave
dissipates through the Langevin thermostat, with the lattice returning to a homogeneous
FM-sign exchange state at long times.
\textbf{Quantitative mesoscale observables.} We extract two quantitative observables
from the simulation trajectory to provide physically testable predictions inaccessible
to direct DFT.
The \textit{domain wall width} $\xi$ characterizes the spatial sharpness of the FM/AFM
boundary. At each constructive interference peak, we identify the dominant AFM cluster,
extract the radial $J$ profile from the cluster centre, and fit it to a hyperbolic
tangent:
\begin{equation}
J(r) = \frac{J_{\rm FM} + J_{\rm AFM}}{2} + \frac{J_{\rm FM} - J_{\rm AFM}}{2}
\tanh\!\left(\frac{r - r_0}{\xi}\right)
\label{eq:tanh_fit}
\end{equation}
where $r_0$ is the domain wall centre position. The fits yield
$\xi = 1.52 \pm 0.52$~nm ($R^2 = 0.84$) at frame~36 and
$\xi = 1.90 \pm 0.48$~nm ($R^2 = 0.91$) at frame~89
(Figs.~\ref{fig:metrics}(b)--(c)), giving a mean domain wall width of
$\xi = 1.7 \pm 0.3$~nm averaged over $N = 2$ constructive interference events. A
notable feature of both radial profiles is that $J$ values at the tensile periphery
reach approximately $+7$--$+8$~meV, significantly exceeding the equilibrium
$J_1 \approx +2$~meV. This reflects the tensile strain at the periphery of the
interference zone, which opens the Cr-I-Cr bond angle and strengthens the FM
superexchange pathway~\cite{Goodenough1955_PhysRev, Kanamori1959_JPCS}, producing a
simultaneous AFM core and enhanced-FM shell within a single strain wave cycle.
The \textit{oscillation period} $\tau$ is measured directly from the AFM Cr fraction
time series $\mathcal{F}(t)$ (Fig.~\ref{fig:metrics}(a)), which exhibits two distinct
constructive interference events at frames 36 and 89, separated by 53 frames (265~fs at
5~fs per frame), yielding $\tau = 0.27$~ps. Because the present framework models
classical structural dynamics without magnetic back-action, this timescale reflects the
wave propagation characteristics of the ML potential rather than a direct prediction for
experimental acoustic phonon periods.
The domain wall width $\xi = 1.7 \pm 0.3$~nm lies within the range accessible to
cryogenic magnetic force microscopy on CrI$_3$ flakes under acoustically induced strain
excitation~\cite{Thiel2019_Science}, providing a direct experimental avenue for
validating the DSpinGNN framework.
\subsection{Internal Consistency of the Physics-Informed Exchange Predictor}
\begin{figure}[htbp]
\includegraphics[scale=0.95]{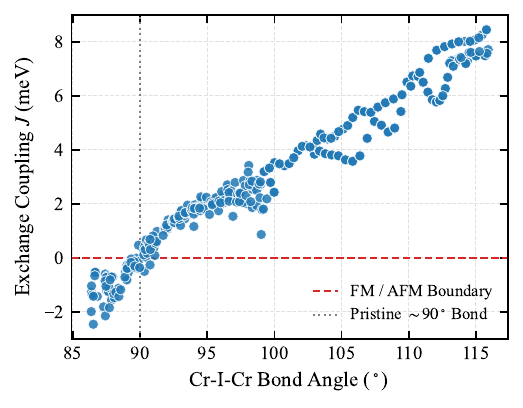}
\caption{\label{fig:kanamori}
Internal consistency check for the physics-informed exchange predictor. Predicted $J$
vs.\ Cr-I-Cr bridging angle $\theta$ sampled from the full 3,200-atom simulation
trajectory. The functional form is consistent with the Goodenough-Kanamori ansatz
(Eq.~\ref{eq:gk_ansatz}), confirming the inductive bias is functioning correctly across
the full mesoscale domain.
}
\end{figure}

As an internal consistency check on the $\Delta$-MLP inductive bias, we examine the
distribution of predicted $J_{ij}$ values as a function of Cr-I-Cr bridging angle
$\theta$ sampled across the full 3,200-atom trajectory (Fig.~\ref{fig:kanamori}). The
predicted $J(\theta)$ relationship follows the functional form of the Goodenough-Kanamori
ansatz (Eq.~\ref{eq:gk_ansatz}), with FM-sign predictions at $\theta > 87^\circ$ and AFM-sign predictions at
$\theta < 87^\circ$, consistent with the GK rules for the competing
superexchange and direct exchange pathways~\cite{Goodenough1955_PhysRev,
Kanamori1959_JPCS, Kashin2020_2DMat}. This confirms the inductive bias is functioning
as designed across the mesoscale simulation domain.
\section{Conclusion}
We have introduced DSpinGNN, a bifurcated equivariant machine learning architecture for
predicting instantaneous isotropic magnetic exchange couplings $J_{ij}$ across a
dynamically deforming crystal lattice. By combining an $E(3)$-equivariant GNN for
force-driven Langevin structural dynamics with a physics-informed $\Delta$-MLP exchange
predictor embedding the Goodenough-Kanamori superexchange relationship, DSpinGNN
achieves near-DFT structural force accuracy and physically constrained exchange
predictions at length scales inaccessible to first-principles methods. On a strictly
withheld test set of 61 configurations---never seen by the model during training or
selection and evaluated exactly once---with energy, force, and exchange coupling assessed
simultaneously against DFT+U ground truth, the model achieves an energy MAE of
1.1~meV/atom, a force MAE of 6.5~meV/\AA, and an exchange coupling MAE of 0.18~meV,
confirming generalization within the collinear, first-nearest-neighbor, isotropic regime.
Deployed in a 3,200-atom ($20\times20$) supercell at 5~K, DSpinGNN maps the exchange
coupling response to a propagating biaxial strain wave. Reflection and superposition of
the wave at periodic boundaries transiently drives localized regions beyond the
DFT-established FM-to-AFM threshold, generating spatially heterogeneous exchange coupling
zones that damp as the wave dissipates. Quantitative analysis yields two mesoscale
observables inaccessible to direct DFT: a domain wall width
$\xi = 1.7 \pm 0.3$~nm extracted via hyperbolic tangent fitting of the radial $J$
profile (mean over $N = 2$ constructive interference events), and a constructive
interference oscillation period $\tau = 0.27$~ps. These constitute testable predictions
for cryogenic magnetic force microscopy under acoustic excitation.
These results are bounded by the stated approximations: collinear Ising constraint,
omission of SOC-driven anisotropic exchange, and adiabatic decoupling of structural and
magnetic branches. Within these constraints, DSpinGNN provides a reproducible, openly
available, length-scale-transferable framework for mapping first-nearest-neighbor
isotropic exchange dynamics in strain-driven 2D magnetic materials.
\section{Limitations and Future Directions}
The current DSpinGNN framework operates within three explicit approximations described in
Sec.~II.B: the collinear Ising constraint, the omission of SOC from the training data,
and the adiabatic decoupled architecture. These jointly restrict the model to the
isotropic exchange regime and exclude magneto-elastic feedback, anisotropic exchange
tensors, and magnon physics. The restriction to first-nearest-neighbor exchange omits
$J_2$ and $J_3$ contributions that influence finer features of the magnetic ground
state~\cite{Lado2017_2DMat}.
Two architectural extensions are natural priorities for future work. First, coupling the
magnetic energy contribution into the force prediction branch would resolve the adiabatic
decoupling and enable true spin-lattice dynamics with back-action of spin state on
phonons. Second, extending the training data to include SOC-derived exchange tensors and
integrating time-reversal equivariant representations~\cite{Yu2023_SpinGNN} would lift
the collinear restriction and allow prediction of the full $3\times3$ exchange tensor
including Kitaev and DMI components---enabling future study of dynamically nucleated
topological spin textures in 2D magnetic materials. Both extensions require substantially
expanded non-collinear DFT training campaigns.
\section*{Data and Code Availability}
All first-principles datasets and automated DFT/Wannier90/TB2J pipelines are publicly
available at~\cite{SpinDFT_repo}. The complete DSpinGNN model architecture, training
framework, and large-scale ASE simulation scripts are available at~\cite{DSpinGNN_repo}.
\begin{acknowledgments}
The authors acknowledge the LUMS High-Performance Computing cluster for DFT data
generation resources. Additional computational resources were provided via RunPod and
DigitalOcean.
\end{acknowledgments}
\appendix
\section{Dataset Composition, Training Protocol, and Model Parameters}
\subsection{Dataset Composition and Splits}
\label{sec:dataset}
The training and validation dataset comprised 406 configurations of monolayer CrI$_3$
under biaxial, uniaxial, and shear strains ($-5\%$ to $+5\%$) with Gaussian atomic
rattling ($\sigma = 0.02$ and $0.04$~\AA). All 467 configurations (including the
subsequent test set) were partitioned using stratified sampling across deformation mode
(biaxial, uniaxial, shear) and strain magnitude bin, ensuring that each split retains
a representative distribution of the full structural deformation space and that no
deformation mode or strain regime is over- or under-represented in any subset. The
resulting splits are a training set (345 configurations) and a validation set (61
configurations) used for early stopping and hyperparameter tuning. After all model
development was finalized, an independent test set of 61 configurations was generated
using the same DFT+U pipeline under equivalent strain and rattling conditions. These 61
configurations were not used at any stage of model training or selection and were
evaluated exactly once, simultaneously assessing energy, force, and exchange coupling
predictions against their DFT+U ground-truth values. The full dataset therefore
comprises 467 configurations in total (345 train / 61 validation / 61 test).
\subsection{Training Protocol}
The E-GNN structural branch was optimized using the Adam
optimizer~\cite{Kingma2014_Adam} (initial learning rate $10^{-3}$, reduced by a factor
of 0.5 on validation plateau with patience 200 epochs) with an MSE loss combining energy
and force terms (force weight: 100). Training ran for 40,000 epochs on an NVIDIA GeForce
RTX~4090 with batch size 64.
The $\Delta$-MLP was optimized using AdamW~\cite{Loshchilov2019_AdamW} (learning rate
$10^{-2}$, weight decay 0.4) with an L1 loss targeting $J_{ij}$ MAE. The analytical
block parameters $\{A, B, C, \alpha, l_{\rm ref}\}$ (Eq.~\ref{eq:gk_ansatz}) are
trainable and optimized jointly with the residual MLP. The six geometric input
features---four Cr-I leg lengths, the Cr-Cr interatomic distance, and the Cr-I-Cr
bridging angle---are encoded via learned CosineSmearing embeddings (32 basis functions
each) for the distances and ChebyshevAngleSmearing (32 basis functions) for the angle,
yielding a concatenated input dimension of $32 \times 6 = 192$. The residual MLP
comprises 2 hidden layers of width 16 with SiLU activations.
\begin{table}[htbp]
\caption{\label{tab:gkparams}
Converged parameters of the Goodenough-Kanamori analytical block
(Eq.~\ref{eq:gk_ansatz}) after end-to-end training.}
\begin{ruledtabular}
\begin{tabular}{lccc}
Parameter & Value & Unit & Physical role \\
\hline
$A$           & $-0.246$      & meV        & Quadratic angle term (cos$^2\theta$) \\
$B$           & $-14.28$      & meV        & Linear angle sensitivity \\
$C$           & $+0.642$      & meV        & FM background at $\theta = 90^\circ$ \\
$\alpha$      & $3.084$       & \AA$^{-1}$ & Bond-length overlap decay \\
$l_{\rm ref}$ & $2.70$        & \AA        & Equilibrium Cr-I bond length \\
\end{tabular}
\end{ruledtabular}
\end{table}
\subsection{Performance Summary}
Final model performance across all splits is reported in Table~\ref{tab:performance}
(main text). The Langevin damping constant used in all mesoscale simulations is
$\gamma = 0.982$~ps$^{-1}$.
\bibliography{manuscript}
\end{document}